# The influence of a power law distribution of cluster size on the light transmission of disordered 1D photonic structures

Michele Bellingeri[1], Francesco Scotognella[2]*
[1] Dipartimento di Fisica, Università di Parma, Viale Usberti 7/A, I-43100, Parma, Italy
[2] Dipartimento di Fisica, Politecnico di Milano, P.zza Leonardo da Vinci 32, I-20133, Milano, Italy
* francesco.scotognella@polimi.it

**Abstract**
A better understanding of the optical properties of random photonic structures is beneficial for many applications, such as random lasing, optical imaging and photovoltaics. Here we investigated the light transmission properties of disordered photonic structures in which the high refractive index layers are aggregated in clusters. We sorted the size of the clusters from a power law distribution tuning the exponent $a$ of the distribution function. The sorted high refractive layer clusters are randomly distributed within the low refractive index layers. We studied the total light transmission, within the photonic band gap of the corresponding periodic crystal, as a function of the exponent in the distribution. We observed that, for $0 \leq a \leq 3.5$, the trend can be fitted with a sigmoidal function.

**Introduction**
Light propagation in dielectric random media offers a variety of fascinating features in the field of diffuse optical imaging [1], random lasing [2-6] and light harvesting for solar devices [7,8]. It is well known that the light transport depends on the dimensionality of the material structure [9]. One-dimensional systems are realized by modulating the dielectric constant in a linear arrangement, as, for example, a random multilayer [3,10,11]. For sake of clarity, a periodic alternation of the dielectric constant results in a one-dimensional photonic crystal [12-15], while a modulation of the dielectric constant that follow a deterministic generation rule gives rise to a quasicrystal [16-20].
In one-dimensional disordered photonic systems clear features related to Anderson localization have been observed [21,22], but also transport phenomena as optical Bloch oscillations and necklaces states [23-26]. Studying the light transmission for different lengths of these systems, oscillations of the average transmission, in a specific spectral range, have been observed as a function of the sample length [27]. Moreover, by grouping high refractive index layers in one-dimensional clusters, randomly distributed within layers of low refractive index, the total transmission, in a specific spectral range, oscillates as a function of the cluster size [28,29].
In this work, we investigated the light transmission properties of disordered photonic structures in which the high refractive index layers are aggregated in clusters. The size of the high refractive index layer clusters is sorted from a family of power law distribution functions. We studied the total light transmission, within the photonic band gap of the periodic crystal, as a function of the power law function exponent shaping the distribution. We discovered that the total light transmission trend as a function of the exponent $a$ of the distribution can be fitted, $0 \leq a \leq 3.5$, with a sigmoidal function.

**Methods**
We realized one dimensional photonic structures made by 360 layers with refractive index $n_1$=1.6 and a thickness $d_1$=70nm, and by 2520 layer with refractive index $n_2$=1.4 and a thickness $d_2$=80nm. Thus, the ratio of high/low refractive index layers in the structure is 1/8.

In a periodic arrangement, resulting in a photonic crystal, the unit cell is composed by a layer with refractive index $n_1$ followed 7 layers with refractive index $n_2$ (360 unit cells in total). Then, we realized a disordered photonic structure where the high refractive index layer ($n_1$) are aggregated in clusters. The probability distribution of the size of the clusters follows a power law function:

$$f(x) = x^{-a} \tag{1}$$

where $x$ is the size of the cluster and $a$ the exponent of the function. The size of the clusters spans from 1 to 50. For this reason, the distribution becomes a truncated power law function. The resulting distribution as a function of the exponent $a$ is shown in Figure 1. For example, a low value of $a$ (e.g. 0.1) results in a structure where small and large clusters have both a significant probability to occur. A high value of $a$ (e.g. 2) results in a structure where clusters made of more than 15 high refractive index layers have a probability to occur that is close to 0. When $a=0$, the probability distribution is the uniform distribution where all the clusters sizes have the same probabilities to occur.

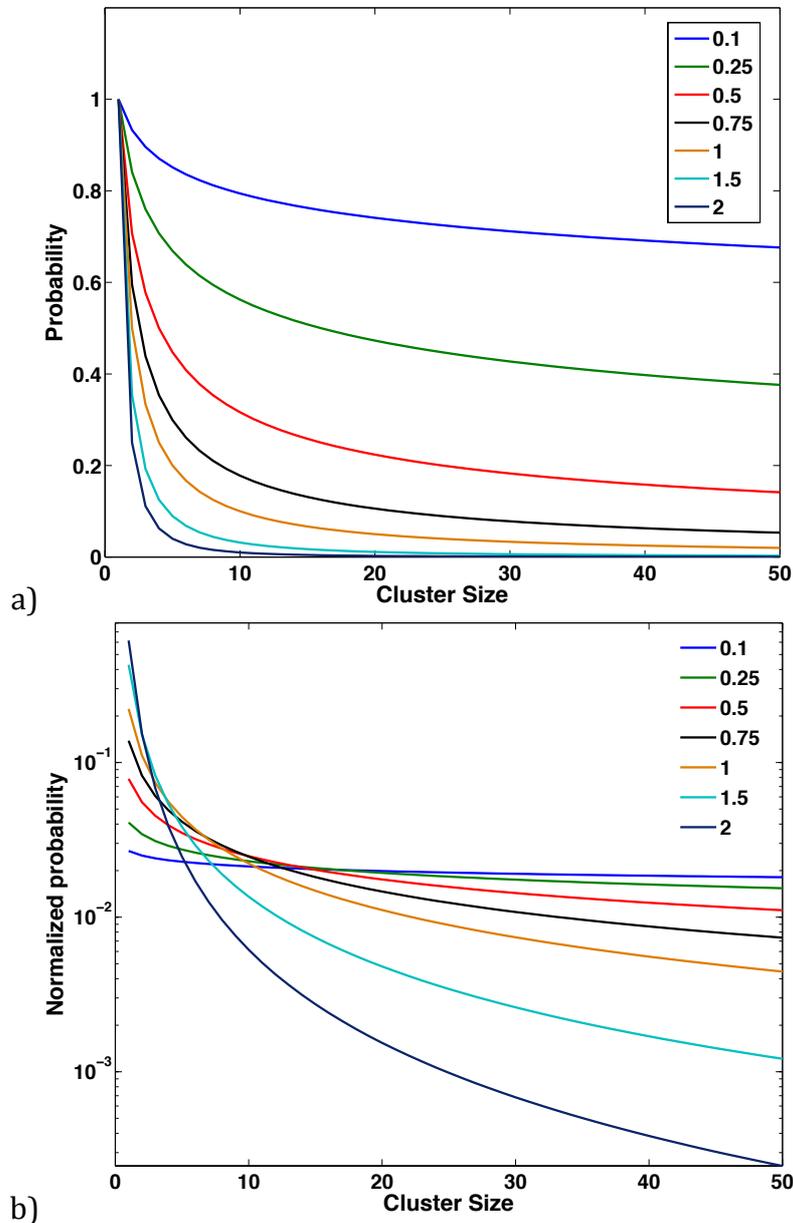

**Figure 1.** a) Examples of power law functions employed for the cluster size distribution in the photonic structures. b) Normalized power law functions

(normalized area under the curve). The exponent *a* of the function $f(x)=x^{-a}$ is reported in the legend.

The clusters of high refractive index layers are randomly distributed within the 2520 low refractive index layers. We have performed the experiment, for each exponent *a*, considering 100 structures characterized by different random cluster arrangements within the low refractive index layers. In this way, we could avoid that the light propagation in the structure depends on a particular arrangement of clusters.

To calculate the transmission spectra of the photonic structures we have used the transfer matrix method [30] for a glass/multilayer/air system (in which glass is the sample substrate, with refractive index $n_s$=1.46). The transmission spectra were calculated as a function of the wavelength with a step of 0.1 nm.

**Results and Discussion**

We calculated the transmission spectrum of the photonic crystal, with 1 high refractive index layer and 7 low refractive index layer in each unit cell (for a total of 360 unit cells), in a spectral range around its fundamental photonic band gap. The crystal shows a photonic band gap with its full width at half maximum (FWHM) between 1763 and 1822 nm (Figure 2, blue curve). The high number of unit cells results in a photonic band gap with negligible light transmission in the range 1764 – 1821 nm.

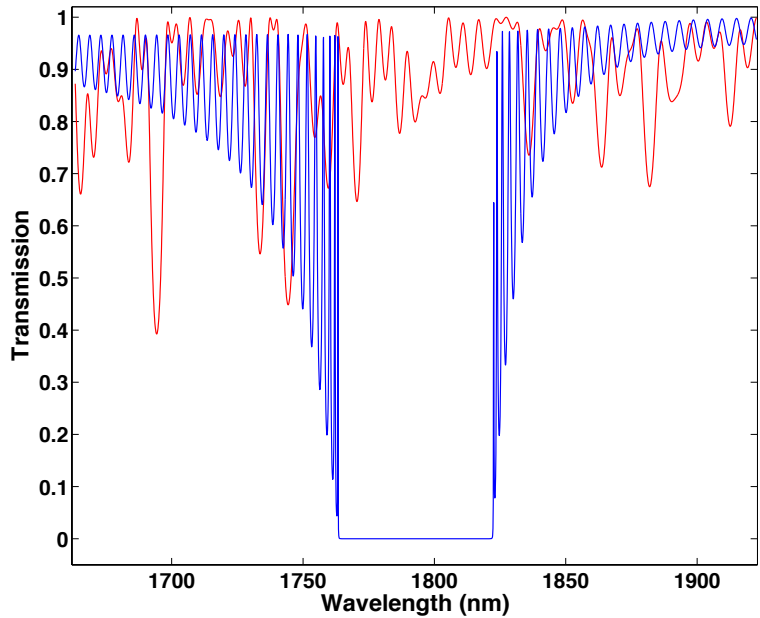

**Figure 2.** Transmission spectra of the periodic crystal (blue curve) and of the disordered structures with a cluster size distribution characterized by a power law with *a*=0.1.

The red curve in Figure 2 corresponds to the transmission spectrum of the disordered structure made with a power law distribution of the high refractive index layer clusters with the exponent *a*=0.1. In the photonic band gap region only weak transmission depths are observed. This behaviour is in agreement with what observed in Ref. [28], where disordered photonic structures characterized by the occurrence of larger clusters show a very weak transmission depths in the spectral region corresponding to the photonic band gap of the periodic crystal. We want to underline that in a photonic structure corresponding to the exponent *a*=0.1 (blue curve in Figure 1) the occurrence of large refractive index layer clusters has a significantly high probability.

In Figure 3 the red curve displays, instead, the transmission spectrum of the disordered structure corresponding to $a$=3.5. The difference between this structure and the one with $a$=0.1 is noteworthy: with the same number of high refractive index layers, the intensity of the transmission depths is much stronger.

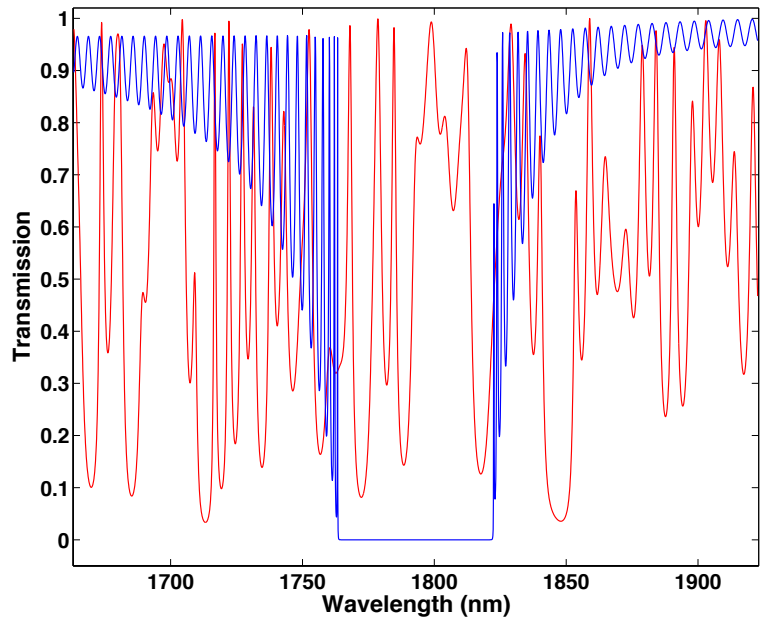

**Figure 3.** Transmission spectra of the periodic crystal (blue curve) and of the disordered structures with a cluster size distribution characterized by a power law with $a$=3.5.

The total transmission in the photonic band gap region (the overall contribution of the transmission between 1763 and 1822 nm, with a step of 0.1 nm) follows a very clear trend as a function of $a$, as displayed in Figure 4. When $a$=0, the distribution probability is the uniform distribution where all cluster sizes present the same occurrence. Increasing $a$, small high refractive index clusters play a major role, and they contribute to the incidence of more intense transmission depths. Between $a$=0 and $a$=2.75 the total transmission decreases more than the 45%.

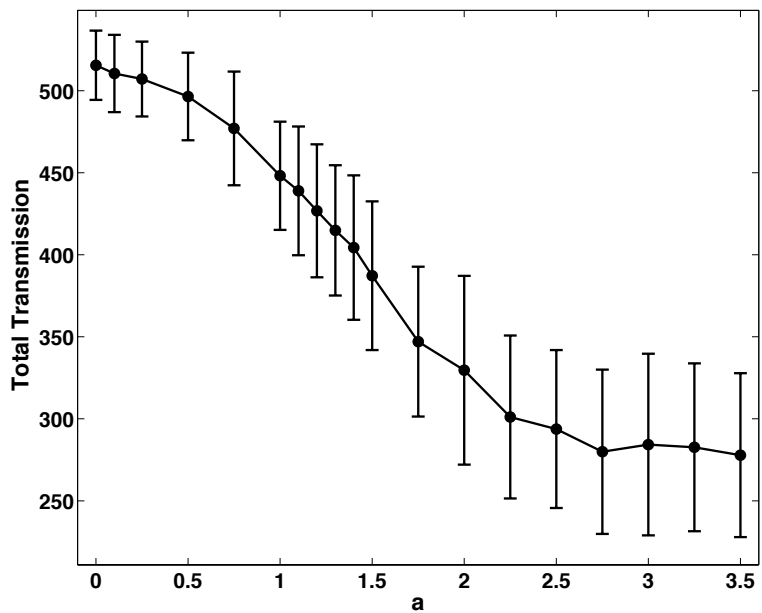

**Figure 4.** Total transmission, calculated in the spectral region corresponding to the periodic crystal band gap (where the photonic crystal show a total transmission equal to 0), of the photonic structures as a function of $a$, exponent in the power law distribution of clusters, up to $a$=3.5. Each point in the curve is an average of 100 values and the error bars are reported.

We observed that the trend of the total transmission can be fit, as shown in Figure 5, with a sigmoidal membership function of $a$ (up to $a$=3.5):

$$f(a) = c_1 - \frac{c_2}{1 + e^{-c_3(a-c_4)}} \quad (2)$$

where $c_1$ is close the maximum total transmission value (523.1), $c_2$ is 247.6 (close to the minimum total transmission value), $c_3$ and $c_4$ are conventional parameters of the sigmoidal function and are, respectively, 2.318 and 1.399 [R-square: 0.9987; adjusted R-square: 0.9984].

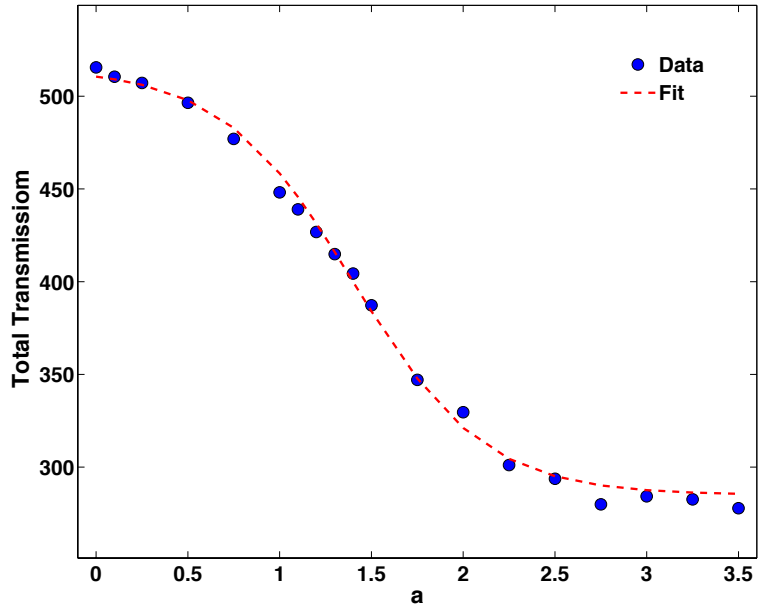

**Figure 5.** Fit of the total transmission data, as a function of the exponent $a$, with a sigmoidal membership function [fit values: R-square: 0.9987; adjusted R-square: 0.9984].

The first derivative of the sigmoidal function, employed to fit the total transmission data, shows a monotonic decrease up to a minimum, corresponding to $a$=1.4, and then an increase. For $a$>3 the first derivative of the function becomes asymptotic to zero. This means that we are converging to a distribution of clusters with size equal to 1 (i.e. single layer). In fact, if we calculate the total transmission of a random sequence of 360 high refractive index layers and 2520 low refractive index layers (in such situation the probability for the high refractive index layers to form clusters is very low), and the average total transmission for 100 permutations is 298.0574 (standard deviation of 49.5189), very close to the total transmission values for $a$>3 (See also Table 1).

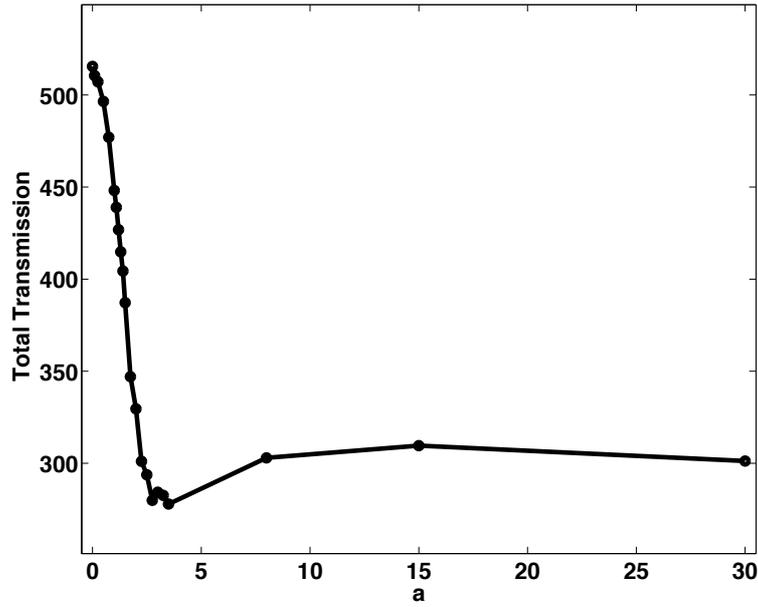

**Figure 6.** Total transmission, calculated in the spectral region corresponding to the periodic crystal band gap, of the photonic structures as a function of $a$, up to $a$=30.

| $a$ | Skewness | Average Total transmission | Standard Deviation |
|---|---|---|---|
| 0 | 0 | 515.5125 | 21.1208 |
| 0.1 | 0.0881 | 510.4924 | 23.5475 |
| 0.25 | 0.2353 | 507.1204 | 22.8104 |
| 0.5 | 0.5349 | 496.4655 | 26.6913 |
| 0.75 | 0.9158 | 477.0094 | 34.6684 |
| 1 | 1.4027 | 448.1437 | 32.9791 |
| 1.1 | 1.6345 | 438.9499 | 39.2207 |
| 1.2 | 1.8919 | 426.7881 | 40.5140 |
| 1.3 | 2.1729 | 414.8488 | 39.6994 |
| 1.4 | 2.4883 | 404.3588 | 44.0218 |
| 1.5 | 2.8339 | 387.2111 | 45.3048 |
| 1.75 | 3.8593 | 347.0453 | 45.6731 |
| 2 | 5.1414 | 329.6135 | 57.5332 |
| 2.25 | 6.6863 | 301.1054 | 49.6896 |
| 2.5 | 8.4281 | 293.7580 | 48.1569 |
| 2.75 | 10.2392 | 279.9232 | 50.0850 |
| 3 | 11.8095 | 284.2965 | 55.3695 |
| 3.25 | 13.1722 | 282.6160 | 51.1722 |
| 3.5 | 13.7419 | 277.8424 | 49.9545 |
| 8 | 18.0110 | 302.9640 | 52.5127 |
| 15 | 181.6904 | 309.6151 | 38.4533 |
| 30 | $10^4$ | 301.2806 | 48.9134 |

**Table 1.** The average of the total light transmission outcomes of the disordered photonic structures as a function of the exponent $a$, with the corresponding standard deviation. In the table the skewness values of the power law function are also reported [31].

The scenario for higher values of *a* is different. As shown in Figure 6, the total transmission stops to decrease for *a*>3.5 and shows a new trend, with a relative maximum at *a*=15. We underline that the total transmission changes for *a*>3.5 are comparable with their standard deviations values (Table 1). For this reason, we avoid to fit such values. It is worth noting that the standard deviation of the total transmission outcomes increases by increasing *a*. This large variation in the outcomes, may be due to the fact that larger clusters, arranged in different ways within the low refractive layer, induce larger fluctuations of the total transmission outcomes.

In Figure 7 we show the total transmission of the disordered photonic structures as a function of the power law skewness values, reported in Table 1. The more the power law function is asymmetric, the less is the total transmission. Although, for skewness larger than 13.7419 (corresponding to *a*=3.5) the total transmission shows a new trend, with a relative maximum, similar to the behaviour shown in Figure 6.

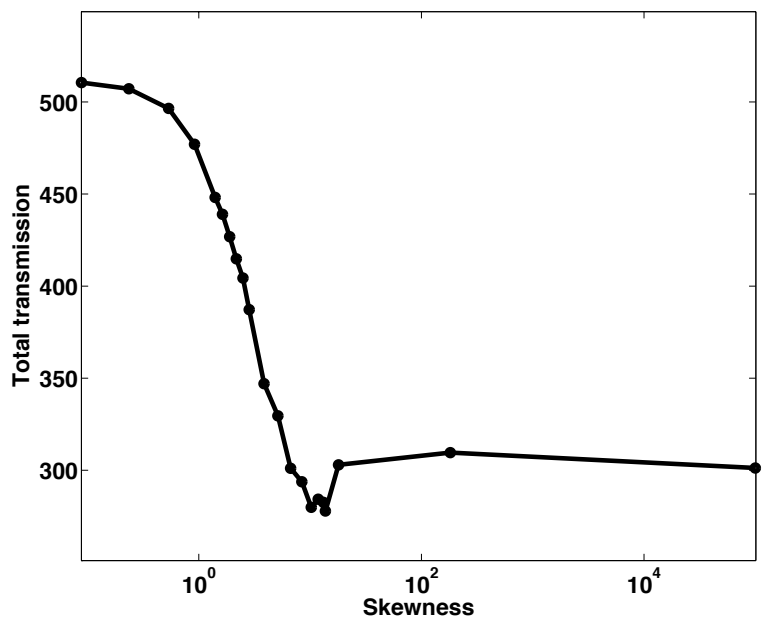

**Figure 7.** Total transmission as a function of the power law function skewness values (reported in Table 1).

In 2013, Fernández-Marín and coworkers studied the electromagnetic transmission through one-dimensional photonic structures in which random layer thicknesses follow a long-tailed distribution, and they found an interesting correlation between the average transmission and the length of the structure [32]. In our work, instead, we study a sample in which the total length of the sample and high/low refractive index layer ratio are the same for all the realizations. Further investigations are needed to better explain the physical phenomenon occurring in this study.

**Conclusions**
In this study we investigated the light transmission properties of disordered photonic structures. In such structures, the high refractive index layers are aggregated in clusters, and the size of the clusters is governed by a power law distribution. The clusters are randomly distributed with the low refractive index layers. We have studied the total light transmission, within the photonic band gap of the corresponding periodic crystal, as a function of the exponent *a* in the distribution. We observed that the trend can be fitted, for 0≤*a*≤3.5, with a sigmoidal function. These results could be interesting for sample with a certain length containing two different materials, e.g. colloidal nanoparticles in a solvent, in which the nanoparticles aggregate with a distribution following a power law.

**Acknowledgements**
The authors would like to thank Dr. Stefano Poletti and Mr. Rubens Camboni for the helpful discussions.

**References**
1. A. P. Gibson, J. C. Hebden, and S. R. Arridge, Phys. Med. Biol. **50**, R1 (2005)
2. H. Cao, Y. G. Zhao, S. T. Ho, E. W. Seelig, Q. H. Wang, and R. P. H. Chang, Phys. Rev. Lett. **82**, 2278 (1999)
3. V. Milner, and A. Z. Genack, Phys. Rev. Lett. **94**, 073901 (2005)
4. R. Frank, A. Lubatsch, and J. Kroha, Phys. Rev. B **73**, 245107 (2006)
5. R. Frank, and A. Lubatsch, Phys. Rev. A **84**, 013814 (2011)
6. S. Lepri, S. Cavalieri, G.-L. Oppo, and D. S. Wiersma, Phys. Rev. A **75**, 063820 (2007)
7. C. Rockstuhl, S. Fahr, K. Bittkau, T. Beckers, R. Carius, F.-J. Haug, T. Söderström, C. Ballif, and F. Lederer, Optics Express **18**, A335 (2010)
8. S. Mokkapati, and K. R. Catchpole, J. Appl. Phys. **112**, 101101 (2012)
9. D. S. Wiersma, Nat. Photonics **7**, 188 (2013)
10. M. Bellingeri, F. Scotognella, Photonic Nanostruct. **10**, 126 (2012)
11. M. Bellingeri, D. Cassi, L. Criante, F. Scotognella, IEEE Photonics J., **5**, 2202811 (2013)
12. K. Sakoda, Optical Properties of Photonic Crystals. Berlin, Germany: Springer-Verlag, 2005.
13. E. Yablonovitch, Phys. Rev. Lett. **58**, 2059 (1987)
14. S. John, Phys. Rev. Lett., **58**, 2486 (1987)
15. J. D. Joannopoulos, R. D. Meade, and J. N. Winn, Photonic Crystals: Molding the Flow of Light. Princeton, NJ, USA: Princeton Univ. Press, 1995.
16. D. Shechtman, I. Blech, D. Gratias, and J. W. Cahn, Phys. Rev. Lett., **53**, 1951 (1984)
17. D. S. Wiersma, R. Sapienza, S. Mujumdar, M. Colocci, M. Ghulinyan, and L. Pavesi, J. Opt. A: Pure Appl. Opt., **7**, S190 (2005)
18. T. Hattori, N. Tsurumachi, S. Kawato, and H. Nakatsuka, Phys. Rev. B, **50**, 4220 (1994)
19. L. Dal Negro, C. J. Oton, Z. Gaburro, L. Pavesi, P. Johnson, A. Lagendijk, R. Righini, M. Colocci, and D. S. Wiersma, Phys. Rev. Lett. **90**, 055501 (2003)
20. Y. H. Cheng, C. H. Chang, C. H. Chen, and W. J. Hsueh, Phys. Rev. A 90, 023830 (2014)
21. J. Topolancik, B. Ilic, and F. Vollmer, Phys. Rev. Lett. **99**, 253901 (2007)
22. Y. Lahini, A. Avidan, F. Pozzi, M. Sorel, R. Morandotti, D. N. Christodoulides, and Y. Siberberg, Phys. Rev. Lett. **100**, 013906 (2008)
23. J. Bertolotti, S. Gottardo, D. S. Wiersma, M. Ghulinyan, and L. Pavesi, Phys. Rev. Lett. **94**, 113903 (2005)
24. P. Sebbah, B. Hu, J. M. Klosner, and A. Z. Genack, Phys. Rev. Lett. **96**, 183902 (2006)
25. J. B: Pendry, J. Phys. C **20**, 733 (1987)
26. A. V. Tartakovskii, M. V. Fistul, M. E. Raikh, and I. M. Ruzin, Sov. Phys. Semicond. **21**, 370 (1987)
27. M. Ghulinyan, Phys. Rev. Lett. **99**, 063905 (2007)
28. M. Bellingeri, F. Scotognella, arXiv:1405.5496v1
29. In Ref. 17 the average transmission is calculated in a narrow range around $c/\lambda_0$. $\lambda_0/4=n_i d_i$, where $n_i$ and $d_i$ is refractive index and the thickness of the layer, and $\lambda_0$=1.5 µm. In Ref. [18] the total transmission has been calculated within the spectral region of the full width at half maximum of the photonic band gap of the corresponding periodic crystal, by integrating over the transmission of the disordered structure within this region. See Ref. 17 and 18 for more details.

30. M. Born, E. Wolf, Principles of Optics, Cambridge University Press, Cambridge, 1999
31. D. N. Joanes, and C. A. Gill, J. Roy. Stat. Soc. Ser. D Statistician, 47, 183 (1998)
32. A. A. Fernández-Marín, J. A. Méndez-Bermúdez, and Victor A. Gopar, Phys. Rev. A **85**, 035803 (2013)